# Stoichiometry, Phase, and Texture Evolution in PLD-Grown Hexagonal Barium Ferrite Films as a Function of Laser Process Parameters

Chengju Yu[1], Alexander S Sokolov[1], Piotr Kulik[1], and Vincent G. Harris[1]

[1]Department of Electrical and Computer Engineering and Center for Microwave Magnetic Materials and Integrated Circuits (CM$^3$IC), Northeastern University, Boston, MA 02115 USA

## Abstract

Barium hexaferrite ($BaFe_{12}O_{19}$ or BaM) films were grown on c-plane sapphire (0001) substrates by pulsed laser deposition (PLD) to evaluate the effects of laser fluence on their composition, structure, and magnetic properties. A Continuum Surelite pulsed 266 nm Nd:YAG laser was employed, and the laser fluence was varied systemically between 1 and 5.7 [J/cm$^2$]. Deviations from the stoichiometric transfer between the BaM targets and deposited thin films occurred as the laser fluence changed. The Fe to Ba ratio in the films increased with laser fluence. The films deposited at laser fluences below 4 J/cm$^2$ showed undesirable 3-dimensional island growth. Moreover, insufficient laser energy resulted in the deposition of some secondary phases, for example, Barium monoferrite ($BaFe_2O_4$) and Magnetite ($Fe_3O_4$). Alternatively, laser fluences above 5 J/cm$^2$ promoted resputtering of the growing film at the substrate and degraded film quality, structure, and magnetic properties. BaM films deposited at 4.8 J/cm$^2$ - the optimal laser fluence - showed excellent c-axis orientation perpendicular to the film plane with an anisotropy field of 16 kOe and saturation magnetization of 4.39 kG. These results clearly demonstrate a strong influence of laser parameters on PLD-grown hexaferrite films and provide a path for high-yield production using pulsed laser depositions systems.

## 1. Introduction

Bulk hexagonal ferrites are employed successfully in nonreciprocal microwave devices, such as phase shifters, isolators and circulators, because of high permeability, non-reciprocal electromagnetic properties, and moderate to low high frequency losses [1]. Compared to their bulk counterparts, ferrite films offer a number of advantages, most importantly the possibility of integration with microwave/millimeter-wave monolithic integrated circuits (MMICs) [2]. Over the past 25 years, several research groups have demonstrated the processing of hexagonal ferrite films with suitable RF properties. Several techniques, such as sputtering [3], sol-gel [4], screen printing [5] and liquid phase epitaxy [6], have been investigated. Compared to said techniques, pulsed laser deposition has higher deposition rates and is a relatively low-cost method with the major advantage of stoichiometric material transfer between target and films if proper deposition conditions are employed. Notwithstanding these advances, this technology remains somewhat immature as a tool for industrial production of useful devices. Currently, there is a strong push to develop consistent and reliable processing protocols for such films, on large substrates, that demonstrate practical MMIC prototype devices.

The goal of developing useful MMIC devices requires a comprehensive understanding of the influence of processing parameters on RF materials performance. While the influence of the process gas (i.e., oxygen) pressure, substrate surface properties, and substrate temperature have been investigated previously [7, 8, 9, 10], the role of laser beam parameters has not been explored in sufficient detail for ferrite materials. Several studies on the influence of laser parameters on the structure and composition of alloy films and simple metallic oxides have been published previously [11, 12, 13]. Presented here is a comprehensive study of the impact of such laser parameters on the epitaxial growth of hexaferrite films.

As mentioned, a Nd:YAG laser was employed, and the laser fluence, beam diameter, and angular distribution were systematically varied. The resulting films were characterized using a multiplicity of techniques, including Vibrating-sample Magnetometer (VSM), X-ray Diffraction (XRD), Scanning Electron Microscope (SEM), Energy-dispersive X-ray Spectroscopy (EDX), and Ferromagnetic Resonance (FMR). It was concluded that laser parameters have a significant impact upon the film's microstructure, composition, and magnetic properties. Other conditions, such as the oxygen pressure and substrate temperature, are meaningful only for invariable laser settings.

## 2. Pulsed Laser Deposition of Barium Hexaferrite Thin Films

PLD inherently allows one to deposit a wide variety of materials, including oxides, nitrides, and carbides. Compared to alternative approaches, PLD offers a number of advantages. Under optimal conditions, the stoichiometry of the deposited films may be very close to the target stoichiometry, and the deposition rate can be as high as 100 nm/min.

PLD processes can be subdivided into three stages: (1) ablation of the solid target by laser, (2) expansion and stabilization of the plasma plume in the PLD chamber and (3) plasma condensation on the substrate. A schematic illustration of the PLD processes is shown in Figure 1.

The interaction of the laser and target primarily results in three sputtering mechanisms: (1) thermal sputtering, (2) electronic sputtering, and (3) hydrodynamic sputtering. The energy of the laser pulses become immediately absorbed by the target and transformed into thermal, electronic excitation, and hydrodynamic energies. The local temperature on the target surface exposed to the laser pulse may reach up to 5000 K in a few nanoseconds. At these high temperatures and heating rates congruent evaporation of all chemical elements in the target occurs. The resulting plume consists of ions, atoms, and splashed particulates. There are three principal mechanisms that contribute to the PLD plume: (1) exfoliation sputtering, (2) surface boiling, and (3) shock-wave recoil pressure expulsion.

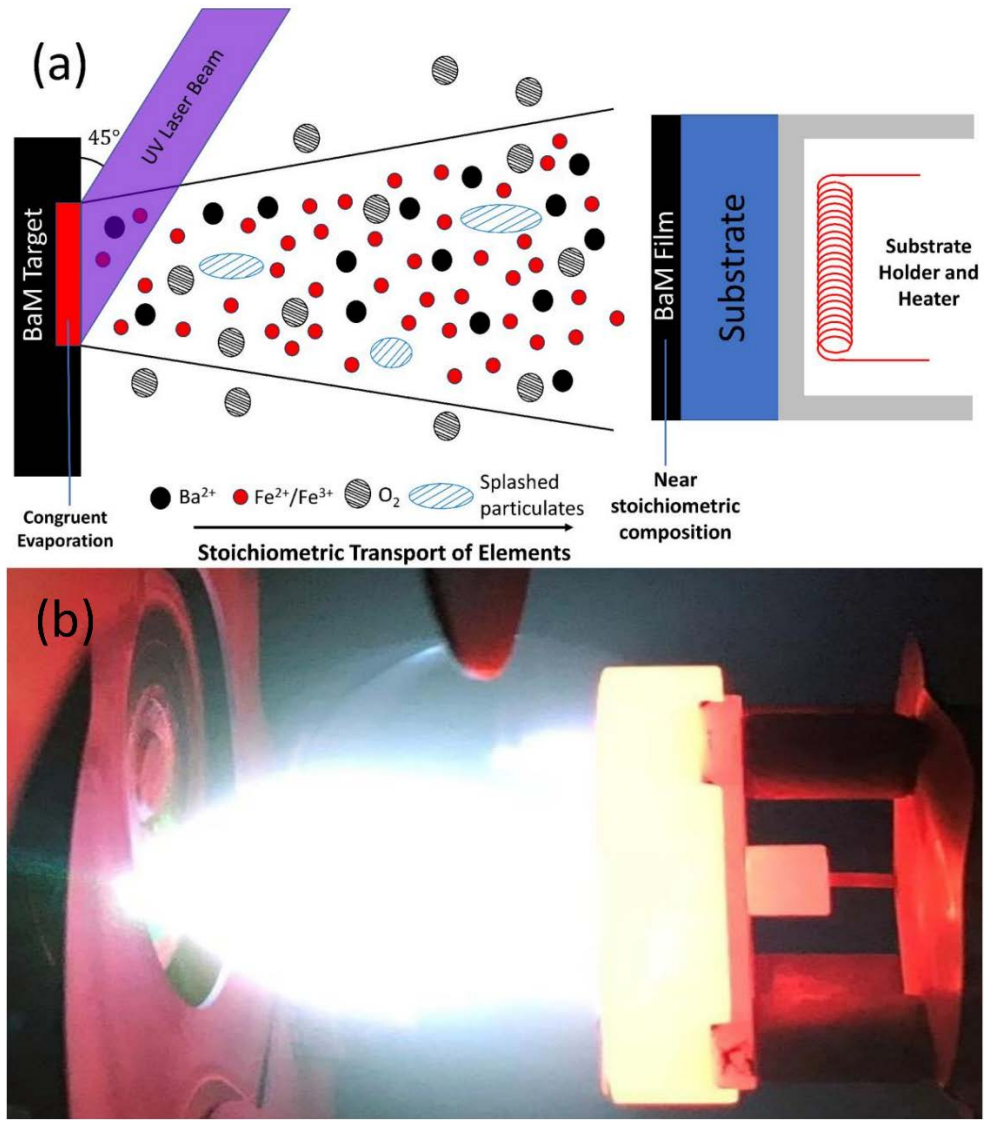

Figure 1. (a) A schematic illustration of the PLD process. (b) Photograph of plasma plume during PLD.

The second stage of the PLD process is the plume expansion. The ions and atoms leave the target with energies as high as 100 eV, but eventually thermalize mainly due to collisions with molecules of the background gas. Therefore, the energy of a given particle depends on its mass and the length of its path prior to collision with the substrate. The resulting element-dependent angular distribution of the energies leads to nonuniform deposition and limits the feasible substrate size. Excessive energies of the arriving particles may also cause element-dependent preferential re-sputtering on the film surface.

It's particularly challenging to grow complex oxide films such as BaM by PLD. The desired outcome can be achieved only under a narrow range of process parameters due to the necessity to transfer several chemical elements with significantly different atomic masses stoichiometrically, and also the complexity of the phase diagram of the Ba-Fe-O system [14] (see Figure 2).

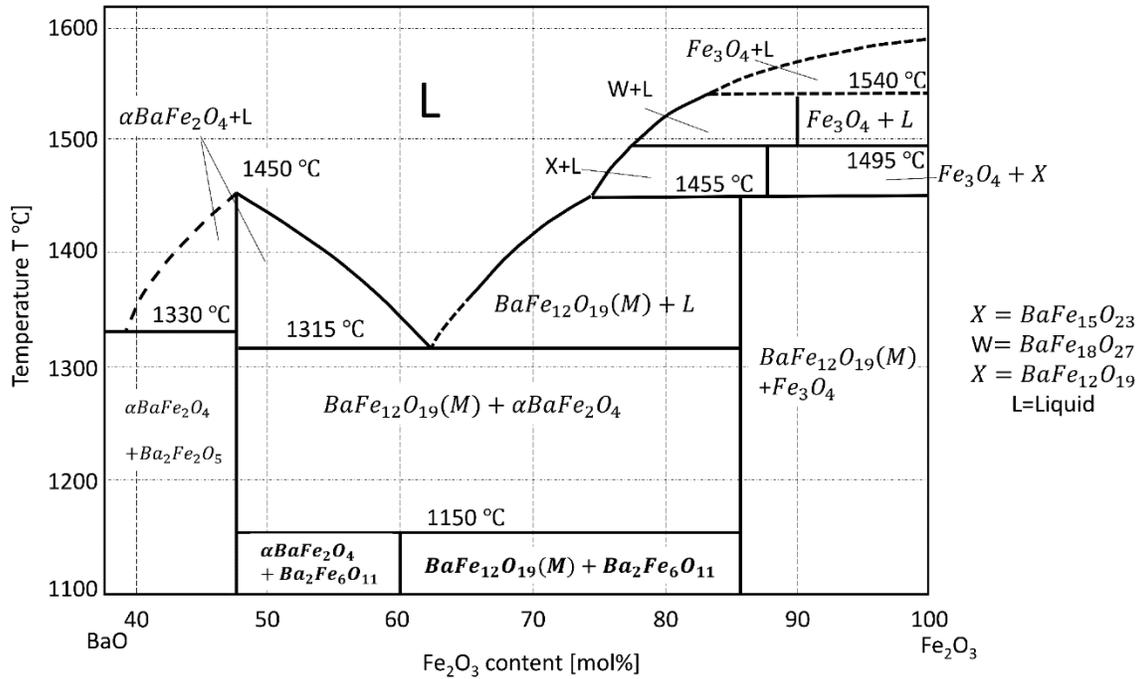

Figure 2. Phase diagram of the Ba-Fe-O system [14].

## 3. Experimental Section

All hexaferrite films for these experiments were grown using the Neocera Pioneer 120 PLD deposition system. A Surrelite solid-state Nd:Yag laser produced by Continuum was used as the source. The laser emitted ten 10 ns pulses per second at 1066 nm. However, the 266 nm harmonic was used for deposition. Before focusing on the target, the beam passed through a series of lenses, mirrors, and a UV transparent chamber entrance window.

A 25 mm BaM target was prepared according to traditional powder metallurgy methods and had a density of near ideal 5.0 g/cm$^3$. The films grown were grown on single crystal polished sapphire

(0001) substrates supplied by MTI Corporation. The substrates were cleaned in acetone, alcohol, and DI-water, and mounted on the heater block using silver paint. The substrate to target distance was set at 70 mm. The laser beam remained fixed, and the target was continuously rotated and rastered in order to achieve a homogeneous ablation. The PLD chamber was evacuated to less than $10^{-6}$ Torr base pressure, and then the substrate was heated to 920 °C. The chamber was then filled with flowing oxygen to maintain a working gas pressure of 200 mTorr during film deposition. Said parameters have been optimized in previous studies and remained unchanged herein [15]. The laser power in this experiment was controlled by the voltage output of the power supply. Figure 3 provides the relation between the applied voltage and the laser energy measured directly in the vacuum chamber. As noted earlier, the power density at the target was intentionally varied between 1 J/cm$^2$ and 5.7 J/cm$^2$. A pre-ablation of the target was achieved by firing the laser for 10 minutes at 10Hz (1000 pulses total) in order to remove the contamination and obtain a steady-state target since less-volatile elements tend to be enrich on the target surface. The total deposition time remained fixed at 1 hour (36,000 laser pulses) for all of films discussed herein. After deposition, the samples were slowly cooled down to room temperature, and the chamber was filled with nitrogen gas. Post-annealing was not employed in this study, as it could obscure some features of the as-deposited BaM film properties.

A number of charactization techniques were employed to analyze the resulting films. The composition and structure of the films were determined by EDXS and XRD (Ultima III Advanced System), respectively. High-resolution sample surface topography images and the film thickness measurements were obtained using a FEI SCIOS SEM system. The magnetic properties presented here were acquired using a Microsense Vibrating Sample Magnetometer (VSM) system.

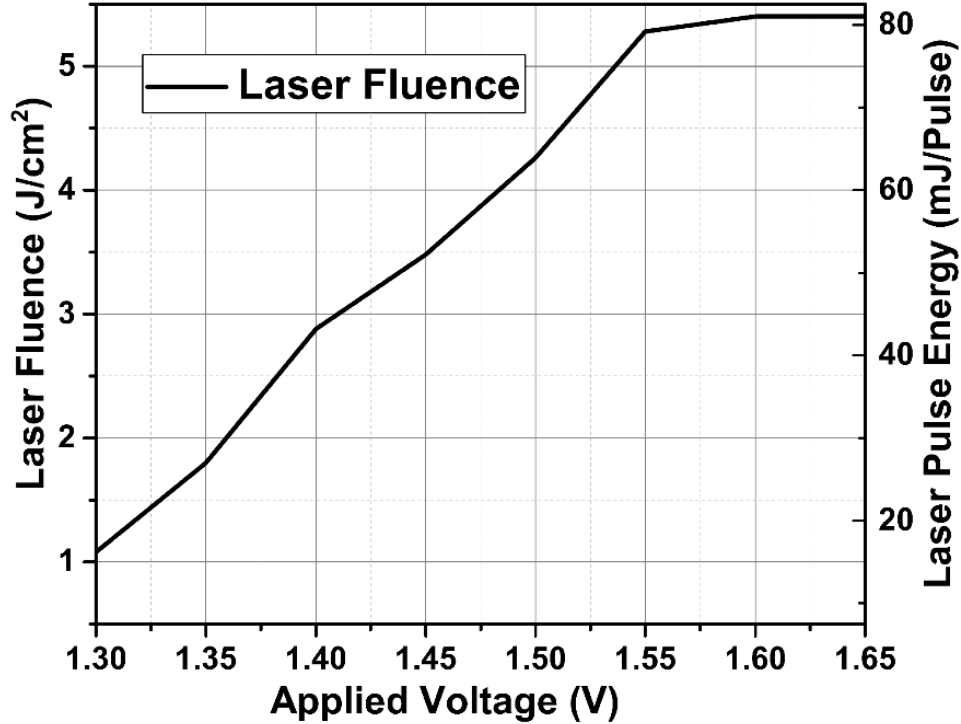

Figure 3. Laser power and fluence as a function of laser power supply voltage.

## 4. Result and discussion

The composition and crystal orientation of the epitaxial films were determined by θ–2θ XRD scans using a Cu-k$_\alpha$ radiation source at room temperature. All peaks were identified using standard XRD patterns for the BaM powder according to the *P63/mmc* space group. A comparison of typical XRD patterns for films grown at 2.5 J/cm$^2$, 4 J/cm$^2$, and 4.8 J/cm$^2$ laser fluences are shown in Figure 4. Significant intensities of the c-axis BaM peaks (0 0 n) reveal the c-textured BaM growth on the substrates at a relatively high laser fluence of 4.8 J/cm$^2$. All films deposited at fluences exceeding 4 J/cm$^2$ showed the c-axis growth of BaM on sapphire substrates. However, non-stoichiometric secondary phases and randomly orientated BaM were observed in the XRD patterns of the films deposited at lower laser fluences (Figure 4). The secondary phases were determined to be the following: barium monoferrite (BaFe$_2$O$_4$), metastable hexagonal ferrite ($Ba_2Fe_6O_{11}$) and magnetite ($Fe_3O_4$). These common secondary phases have been also detected in a number of previous studies [16, 17, 18, 19]. Unlike the samples deposited at the optimal laser power density (4.8 J/cm$^2$), said films contained randomly oriented BaM grains.

Energy-dispersive X-ray spectroscopy (EDX) was utilized to characterize the chemical composition of the PLD-grown BaM films. Nonstoichiometric material transfer in the pulsed laser deposition system was clearly observed. It was determined that the Fe to Ba ratio in the films increases with the laser fluence. Figure 5 shows the Fe/Ba ratio of the films grown under various deposition conditions. Previous studies reported that a congruent ablation happened above some certain laser fluence threshold [20 and 21]. In this study, the Fe/Ba ratio increased continuously and there was no obvious critical point. It can be attributed to the much lower energy (as compared to excimer lasers) of the solid-state laser. Several mechanisms may be considered in order to explain the non-stoichiometric transfer observed during the PLD depositions. 1. Preferential ablation of certain chemical elements from the surface of the target. 2. Element-dependent thermalization/scattering of the plasma plume. 3. Preferential resputtering of certain elements on the surface of the film. As noted earlier, the Fe/Ba ratio dramatically increased with the laser fluence in this study.

Let's first consider the preferential ablation of certain chemical elements from the surface of the target. This mechanism should not play a major role here as the lowest laser fluence in this study was sufficiently high to achieve congruent ablation.

The element-dependent thermalization of the plasma plume, on the other hand, must be the primary mechanism that determines the Fe/Ba ratio in the PLD-grown films. Given that the atomic mass of Ba exceeds the one of Fe by more than a factor of two, Fe atoms and ions lose their energy and scatter much faster due to collisions with the background oxygen gas. Therefore, if the laser fluence is insufficient, fewer Fe atoms reach the substrate, resulting in Ba-rich films.

As the energy of the ions and atoms reaching the substrate increases with the laser fluence, the preferential re-sputtering on the film surface becomes more likely [22, 23]. Let's consider a sputtering process with two independent targets ((1) Ba and (2) Fe). The ratio between the two elements can be described by the following equation

$$\frac{Y_1}{Y_2} = \frac{C_1}{C_2}(\frac{M_2}{M_1})^{2m}(\frac{U_2}{U_1})^{1-2m}$$

where C is the concentration, M is the mass and U is the cohesive energy, and m is the mass effect coefficient (roughly 0.05~0.1 [24, 25]). The ratio $\frac{Y_1}{Y_2}$ then gives the potential of resputtering for these two elements. Considering that the cohesive energies for barium and Iron are 1.9 eV and 4.28 eV respectively, the equation predicts preferential re-sputtering of barium and Fe-rich films at higher laser fluences. Therefore, the laser output should be controlled precisely in order to keep the Fe/Ba ratio in agreement with the stoichiometry of BaFe$_{12}$O$_{19}$.

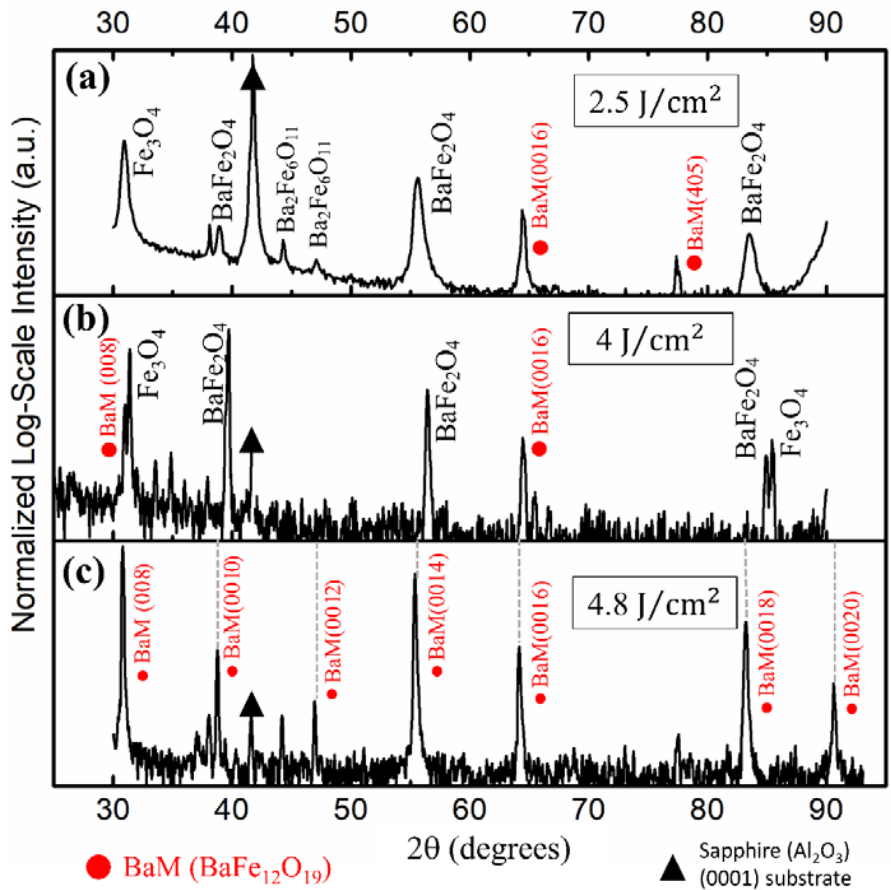

Figure 4. XRD Spectra of BaM films on sapphire substrates deposited at the following laser fluences: (a) 2.5 J/cm², (b) 4 J/cm², (c) 4.8 J/cm².

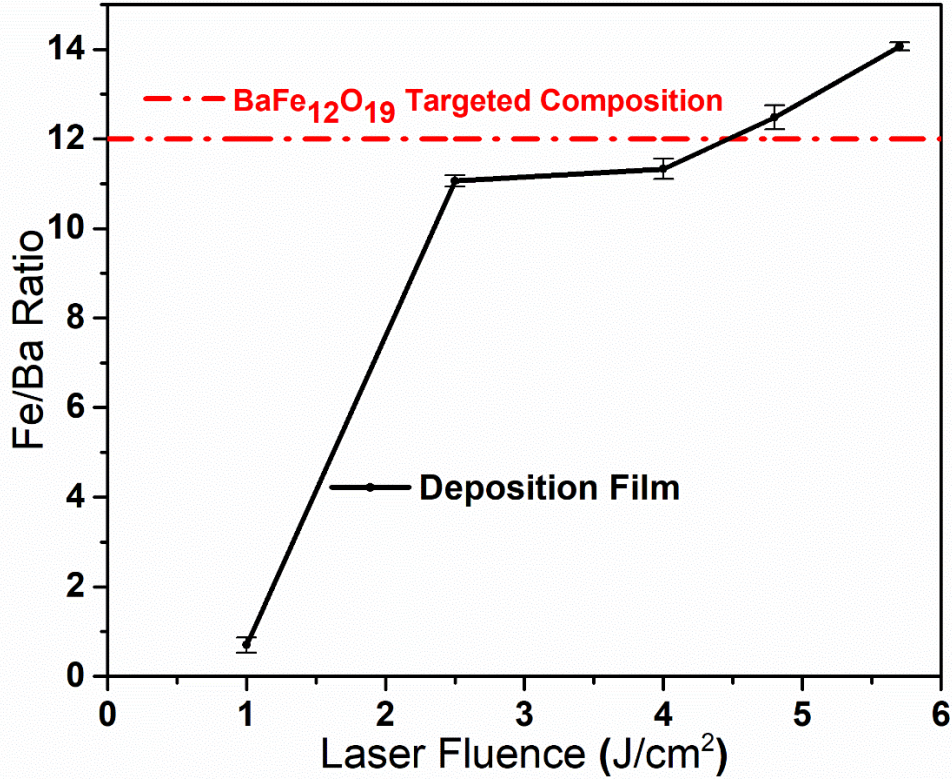

Figure 5. The Fe/Ba ratio in the ferrite films as a function of the laser fluence measured by EDX.

High resolution surface micrographs and cross-section images of the samples shown in Figures 6 were obtained using a FEI SCIOS SEM. The figures demonstrate the evolution of the microstructures and morphologies of the films surfaces as functions of the fluence. The samples deposited at higher laser fluences had relatively smooth surfaces. Rough surfaces were typical for samples deposited at lower power conditions. Films grown at relatively low laser fluences of 2.5J/cm² and 4J/cm² had small grains, velar grain boundaries, and rough surfaces. This indicates that the three-dimensional (3D) island growth mode dominated the process. As depicted in Figure 6 (d), the smoothest film surfaces were achieved at 4.8 J/cm². This implies that the films mostly experienced step flow growth. Generally, the growth mode is primarily determined by the dynamics of the surface energies and the plasma energy. Given that the working gas pressure (Oxygen) and the substrate temperature were fixed in this study, only the plume plasma energy varied as a function of laser fluence. Figure 6 (a) shows the surface morphology of an area close to the edge of the film deposited at 4.8 J/cm². Uniform hexagonal grains are clearly exposed, and the crystallite size was measured to be around 0.4~0.5 µm, while the edges of the hexagonal grains were in almost perfectly registry. When the laser fluence was too high (i.e., 5.7 J/cm$^2$, Figure 6 (e)), the film surface once again became rougher and magnetic coercivity of the films increased due to effective surface pinning of domain wall motion. The grain outgrowth induces lattice distortions, grain cracks, and defects, and eventually affects the magnetic properties of the films.

Some elongated grain edges can be seen piercing the surfaces of the highly textured films deposited at laser fluences in the range between 4.8 J/cm$^2$ and 5.7 J/cm$^2$. Similar BaM film surface features have been reported in earlier studies [26,27]. These edges belong to c-axis oriented grains on the surface. Said grains align along the film plane and indicate that the film growth beneath also preferentially happened along the crystallographic c-plane. Despite some imperfections on the surfaces of the films deposited at higher fluences, no cracks were observed.

Focused ion beam (FIB) milling (Ga+ primary ion beam) was employed to measure the thickness and observe the cross-sections of the BaM films (Figure 6 (f)). Prior to FIB milling, a Pt-based protection layer is deposited above the interested area by means of EBID (electron-beam induced deposition) in order to protect the surface from incurring FIB induced damage unwanted milling. It was determined that the deposition rate strongly correlated with laser fluence, given that the spot size remained unchanged. The rate grew from 0.06 mm/hour at 2.5 J/cm$^2$ to 0.3 mm/hour at 4 J/cm$^2$ and to 0.4 mm/hour at 4.8 J/cm$^2$.

The plume energies influence strongly PLD film growth. The following experiment was designed to evaluate the spatial distribution of the film parameters on the substrate surface. A 10 mm square sapphire substrate was intentionally mounted with an offset with respect to the plume. The resulting film on this substrate had two distinct areas with different appearances (Figure 7). The side of the film that was closer to the center of the plume appeared as a high quality BaM film grown at the optimal laser fluence – the fluence used in this experiment. The outer side of the film looked similar to the films grown using insufficient laser fluences. Both EDX and SEM were used to characterize these two areas. The Fe/Ba ratio gradually decreased towards the center of the plume, from 15.8 to 12.7. This observation revealed a non-stoichiometric angular distribution of chemical elements in the plume, a phenomenon that has been observed in other studies [20, 28].

The Fe-rich area closer to the outer edge of the plume is consistent with the fact that the lighter Fe ions and atoms should have a broader angular distribution than the heavier Ba species. Figure 7 illustrates that the surface morphologies of these two distinct areas of the film differ dramatically. The area closer to the center of the plume appears to have a high density and high degree of crystalline orientation – similar to what was observed in other high quality BaM films. In contrast, the outer edge has a granular structure and noticeable porosity.

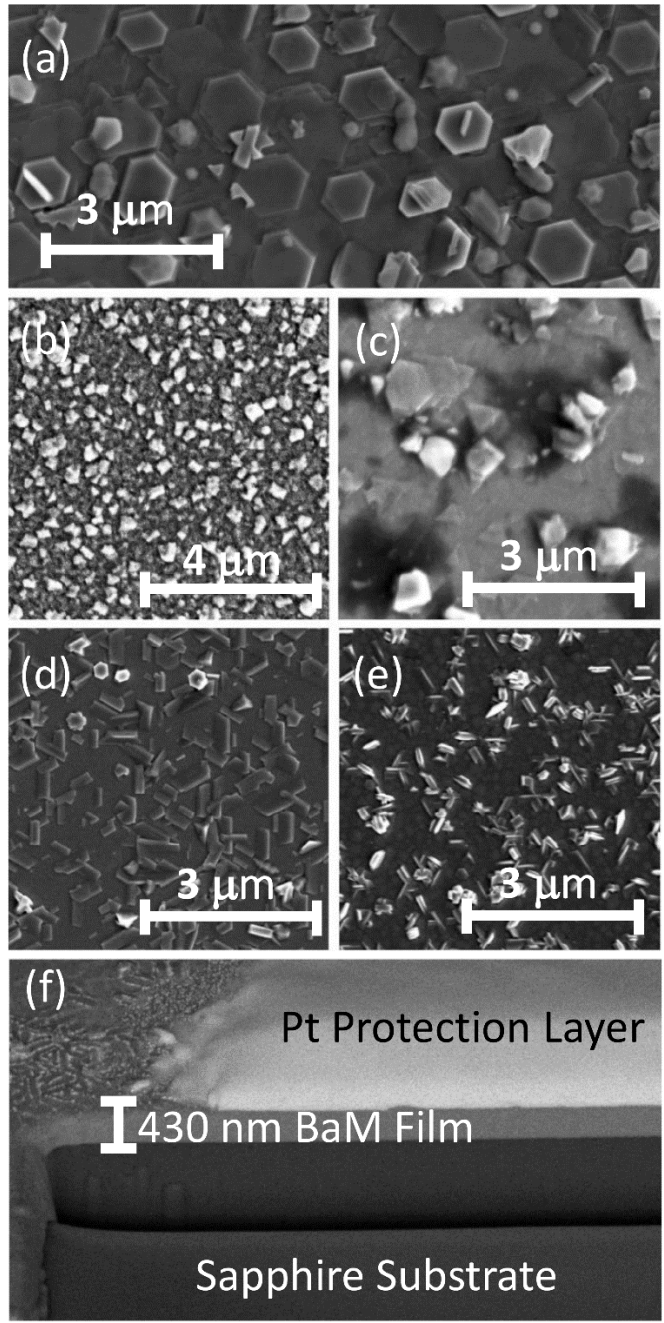

Figure 6. SEM images of the surface morphology of BaM films deposited at (a) 4.8 J/cm², (b) 2.5 J/cm², (c) 4 J/cm², (d) 4.8 J/cm², and (e) 5.7 J/cm². (f) SEM image of a BaM film cross-section deposited at 4.8 J/cm² laser fluence.

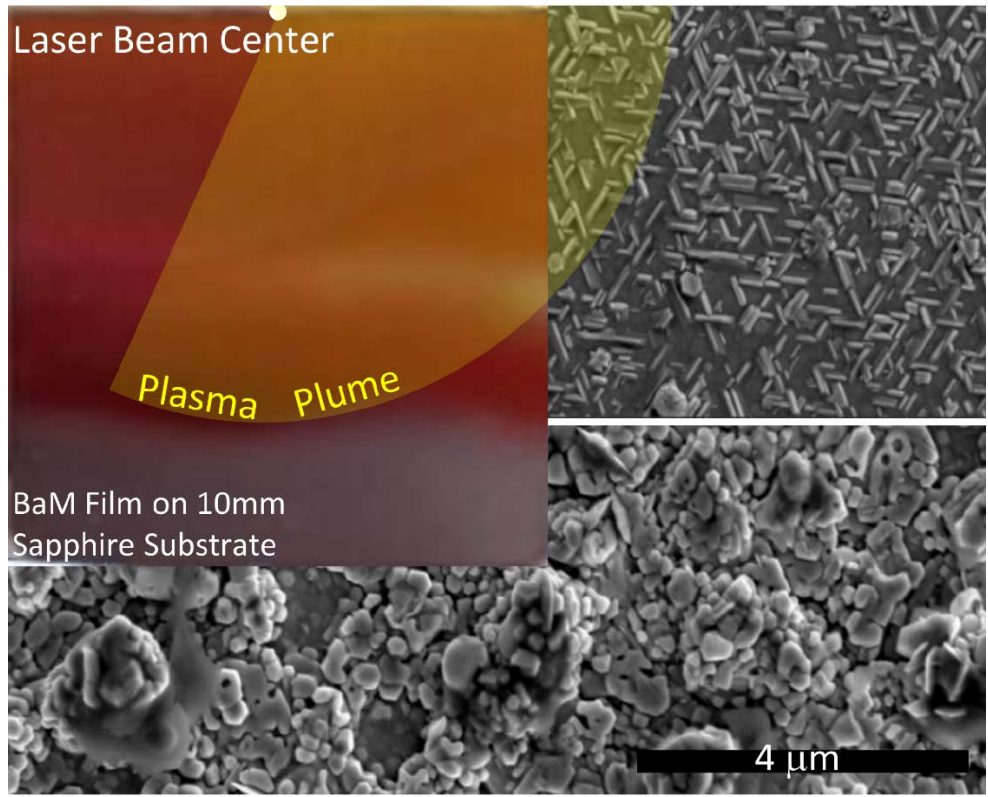

Figure 7. SEM images of different regions of a BaM film deposited on a 10 mm sapphire substrate mounted with an offset with respect to the plume.

Finally, the static and dynamic magnetic properties of the PLD films were measured using VSM and FMR. In-plane and out-of-plane hysteresis loops of the films deposited at different laser fluences are shown in Figure 8. The applied dc magnetic field varied form -20 kOe to +20 kOe in all of these data. In agreement with the XRD measurements, the static magnetic properties of the film deposited at a low fluence of 2.5 J/cm$^2$ are nearly isotropic (Figure 8 (a)). Non-stoichiometric deposition at lower fluences resulted in secondary phases that disrupt the epitaxial growth of BaM. Figure 8 (b) clearly shows a significant difference between the in-plane and out-of-plane hysteresis loops and indicates a significant degree of crystallographic alignment in the film grown at 4.8 J/cm$^2$. The saturation magnetization of said film equaled 4.39 kG. This value is very close to the theoretical saturation magnetization for bulk BaM - 4.46 kG. The magnetic anisotropy field was evaluated using the singular point detection (SPD) method and equaled 16 kOe -again, just slightly lower than the 17 kOe value that has been previously reported for highly textured bulk BaM samples [29]. These lower values may be attributed to the substrate-induced strain arising from lattice and thermal expansion coefficient mismatches and surface imperfections. It's worth noting that the saturation magnetization and magnetic anisotropy can be further increased by post-annealing.

Compared to the film deposited at 4.8 J/cm$^2$, the film deposited at 5.7 J/cm$^2$ once again demonstrated markedly different static magnetic properties (Figure 8 (c)). The coercivity increased from 920 Oe to 1500 Oe and saturation magnetization decreased from 4.36 kG to 4.02 kG. The result is in agreement with other measurements for this sample. This film was Fe-rich (Fe/Ba≈14) and therefore contained secondary phases. There was also observed a number of defects on the surface.

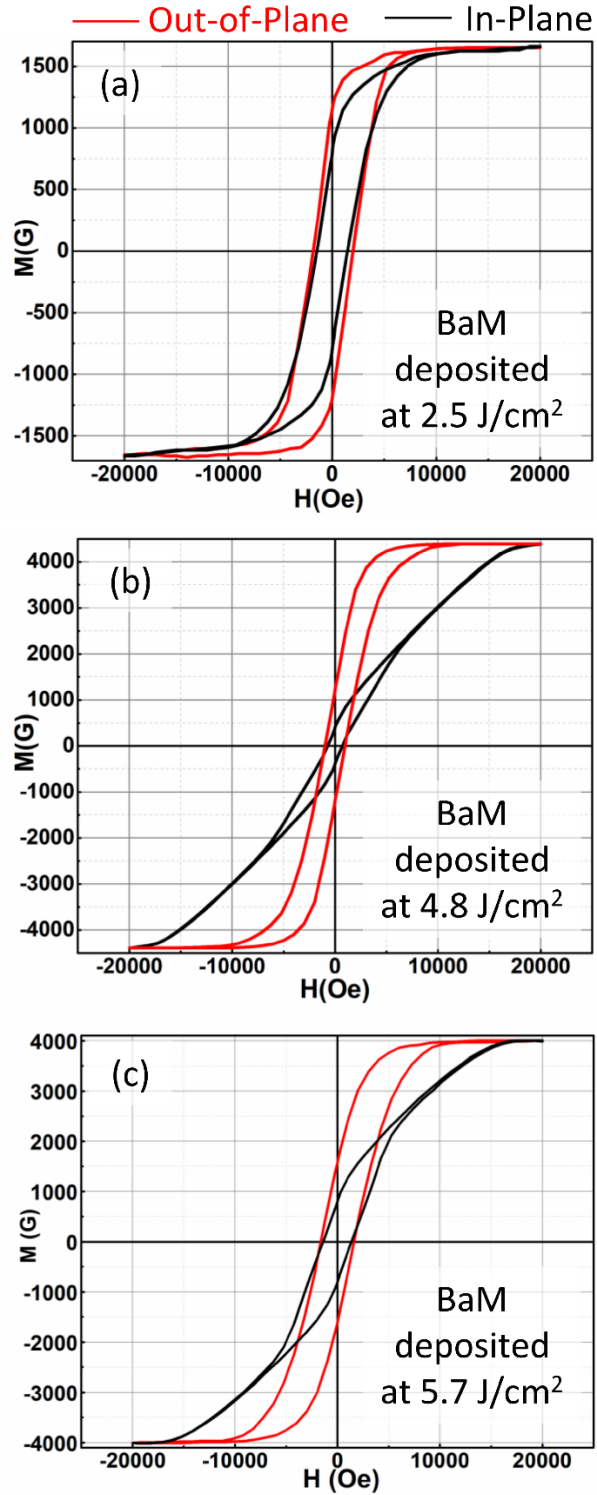

Figure 8. In-plane and out-of-plane hysteresis loops of BaM films PLD grown on (0001) sapphire substrates deposited at (a) 2.5 J/cm$^2$, (b) 4.8 J/cm$^2$, and (c) 5.7 J/cm$^2$.

Dynamic magnetic properties of the films were measured using a coplanar waveguide reflection resonator [30,31]. The actual FMR measurement setup and FMR data are shown in Figure 9.

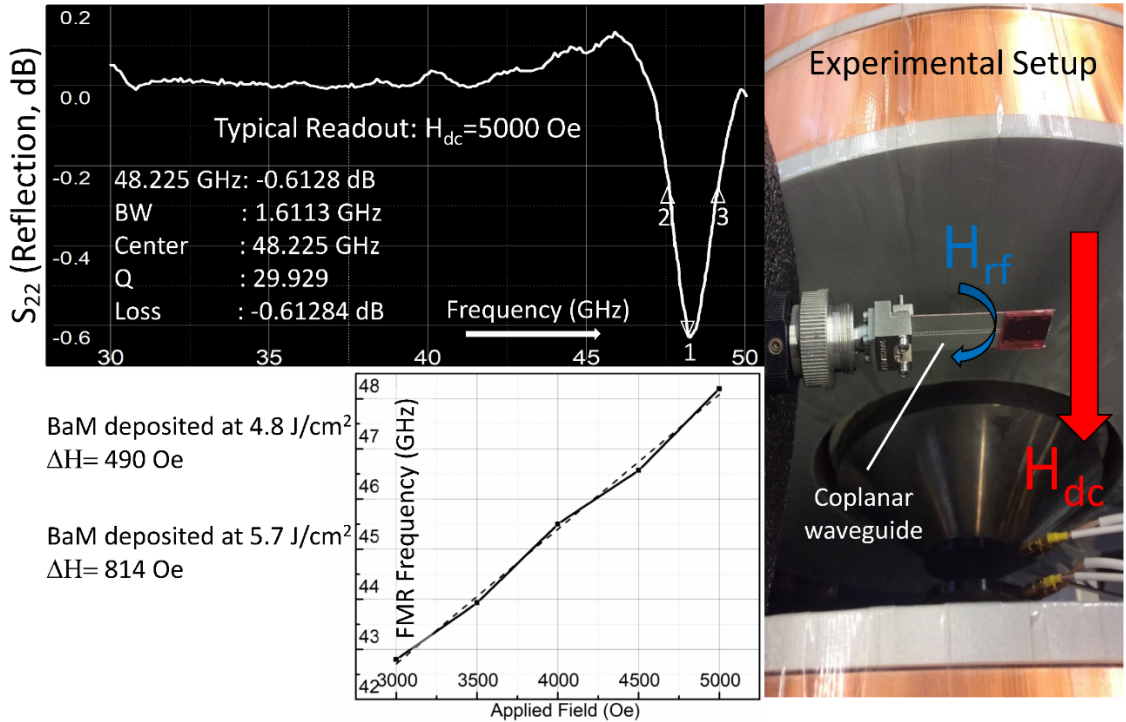

Figure 9. Ferromagnetic resonance measurement setup and results.

In agreement with all other characterization results, the best (i.e., the lowest) FMR linewidth of 490 Oe was measured for the film deposited at 4.8 J/cm². As with the coercivity, this value can be further decreased by employing post annealing in an oxygen gas environment. This step is intended to reduce the strain, eliminate defects, and increase the resistivity of most oxide films. The films deposited at sub-optimal laser fluences demonstrated much broader linewidths. Additionally, the FMR center frequency was measured as a function of the applied dc magnetic field. The relationship between these values provide an alternative way to evaluate anisotropy field and saturation magnetization using the following equation: $f_{FMR} = \gamma[H_{dc} + H_a - (4\pi M_s \cdot N)]$ [32].

# 5. Conclusion

In summary, this study demonstrated that high quality epitaxial hexaferrite films can be grown by pulsed laser deposition only under a narrow range of process parameters. The laser fluence has a significant influence on the stoichiometry, phase, crystallographic texture, and magnetic properties of the deposited films. Relatively low laser fluences are not able to transfer stoichiometric compositions between the target and substrate. Alternatively, excessive laser fluences increase the likelihood of secondary sputtering from the film surface. The Fe/Ba ratio in the films significantly increases with laser fluence but must remain fixed at the correct value in order to produce pure phase BaM. Therefore, nonoptimal laser fluences result in secondary phases, poor epitaxy, and defects. Eventually, nonoptimal laser fluences adversely affect static and dynamic magnetic properties of the films. Finally, the optimal laser fluence can only be determined for a fixed set of other PLD process parameters.

# Acknowledgement

This work was supported by the U.S. Army under Grant W911NF-17-C-0029.